\RequirePackage{fix-cm}
\documentclass[aps,pra,twocolumn,showpacs,superscriptaddress,10pt,longbibliography,floatfix]{revtex4-2}
\usepackage{multirow}
\usepackage{amsmath}
\usepackage{amssymb}
\usepackage{graphicx}% Include figure files
\usepackage{graphics}
\usepackage{dcolumn}% Align table columns on decimal point
\usepackage{bm}% bold math
\usepackage{xcolor}

\usepackage{mathptmx}
\usepackage{etoolbox}
\usepackage{hyperref}
\newcommand{\scs}{\scriptscriptstyle}
\newcommand{\smallm}{\scs (\!-\!)}
\newcommand{\smallp}{\scs (\!+\!)}

\newcommand{\sml}{\scs{\leftarrow}}
\newcommand{\smr}{\scs{\rightarrow}}
\graphicspath{{Fig/}} % Figures pathway

\begin{document}

\title{Uniguided and unicoupled resonances in an asymmetric acoustic metasurface}

\author{Andrey S. Konchakov}
\affiliation{School of Physics and Engineering, ITMO University, 197101, St. Petersburg, Russia}

\author{Daniil S. Buzin}
\affiliation{School of Physics and Engineering, ITMO University, 197101, St. Petersburg, Russia}

\author{Dmitrii N. Maksimov}
\affiliation{Kirensky Institute of Physics, 660036, Krasnoyarsk, Russia}
\affiliation{IRC SQC, Siberian Federal University, 660041, Krasnoyarsk, Russia}

\author{Mingzhao Song}
\affiliation{Qingdao Innovation and Development Center, Harbin Engineering University, Qingdao 266000, Shandong, China}

\

\author{Y.~Li}
\affiliation{School of Physics and Engineering, ITMO University, 197101, St. Petersburg, Russia}
\affiliation{Institute of Acoustics, Tongji University, 200092, Shanghai, China}

\author{Mikhail I. Petrov}
\affiliation{School of Physics and Engineering, ITMO University, 197101, St. Petersburg, Russia}

\author{Chao Peng}
\affiliation{State Key Laboratory of Photonics and Communications, School of Electronics, \& Frontiers Science Center for Nano-optoelectronics, Peking University, Beijing 100871, China}

\author{Andrey A. Bogdanov}
\email{a.bogdanov@hrbeu.edu.cn}
\affiliation{School of Physics and Engineering, ITMO University, 197101, St. Petersburg, Russia}
\affiliation{Qingdao Innovation and Development Center, Harbin Engineering University, Qingdao 266000, Shandong, China}

\author{Pavel S. Pankin}
\email{pankin@itmo.ru}
\affiliation{School of Physics and Engineering, ITMO University, 197101, St. Petersburg, Russia}

\date{\today}

\begin{abstract}
We theoretically study an acoustic metasurface that exhibits ultimate asymmetry in absorption under oblique incidence. This behavior arises from a counter-propagating pair of uniguided (UGR) and unicoupled (UCR) resonances.
The UGR radiates to only one side, while the UCR radiates to both sides but can be excited from only one. In the lossless regime, the transmission phase winds by $2\pi$ across both UCR and UGR while the transmittance remains flat at any prescribed nonzero level.
Thermoviscous losses introduce resonant intensity modulation and enable critical coupling, yielding perfect sound absorption exclusively for incidence through the channel that excites the UCR.
Therefore, asymmetric metasurfaces with UCR and UGR offer a route to full phase control and perfect sound absorption. 
\end{abstract}

\maketitle
%%%%%%%%%%--INTRODUCTION---%%%%%%%%%%%%%%%%%%%
%\tableofcontents

Eigenmodes of open non-Hermitian systems are generally radiative, with complex eigenfrequencies $\omega_r = \omega_0 - i\gamma$, under the temporal sign convention $e^{-i \omega t}$.   
These modes, also known as quasi-normal modes or resonances, are spatially divergent under outgoing-wave boundary conditions and generally non-orthogonal under the usual inner product, which makes them difficult to normalize and analyze
theoretically~\cite{sauvan2022normalization}.
Despite this fact, resonances play a central role in wave systems, mediating coupling between an open wave system and its environment.
Among modern wave systems, metasurfaces, or subwavelength gratings, are distinguished by their fabrication simplicity and the ability to engineer resonances.
Resonances can be adjusted by tailoring the symmetry of metasurface elements or by varying the materials used for metasurface fabrication~\cite{kuznetsov2024roadmap}.

The eigenmodes in metasurfaces include bound states in the continuum (BICs), uniguided
resonances (UGRs) and unicoupled resonances (UCRs).
A BIC is a nonradiative eigenmode of a metasurface whose eigenfrequency $\omega_{BIC} = \omega_0$ is real in the absence of absorption and lies in the radiation continuum~\cite{wang2024optical}.
UGRs can be viewed as semi-BICs because they do not radiate in one direction, either to the
substrate or the superstrate~\cite{yin2020observation, mukherjee2021unidirectional, delplace2025multimodal}.
They have been used for unidirectional couplers~\cite{wang2025low, wang2024ultralow}, lasers~\cite{fang2025abnormal}, emitters~\cite{choi2025unidirectional,cheng2026dynamic}, enhanced Goos-H\"anchen shifts~\cite{wu2025controlling}, tunable reflectors~\cite{hu2025analysis}, interferometers~\cite{Xu2023}, perfect absorbers~\cite{zheng2025dynamical, Xu2023a}, polarization converters~\cite{han2025ultra},  and intensity-flattened phase shifting \cite{Zhang2021a, Zhang2022, hidalgo2025cloaked}.
Transformations between BICs and UGRs are governed by their nontrivial topological properties in parameter space~\cite{Jiang2023, Yin2023, Zhuang2024, wang2025topological, huang2025evolution, zhou2025rotation, choi2025topological}.

\begin{figure}[tbp]
    \centering
    \includegraphics[width=\linewidth]{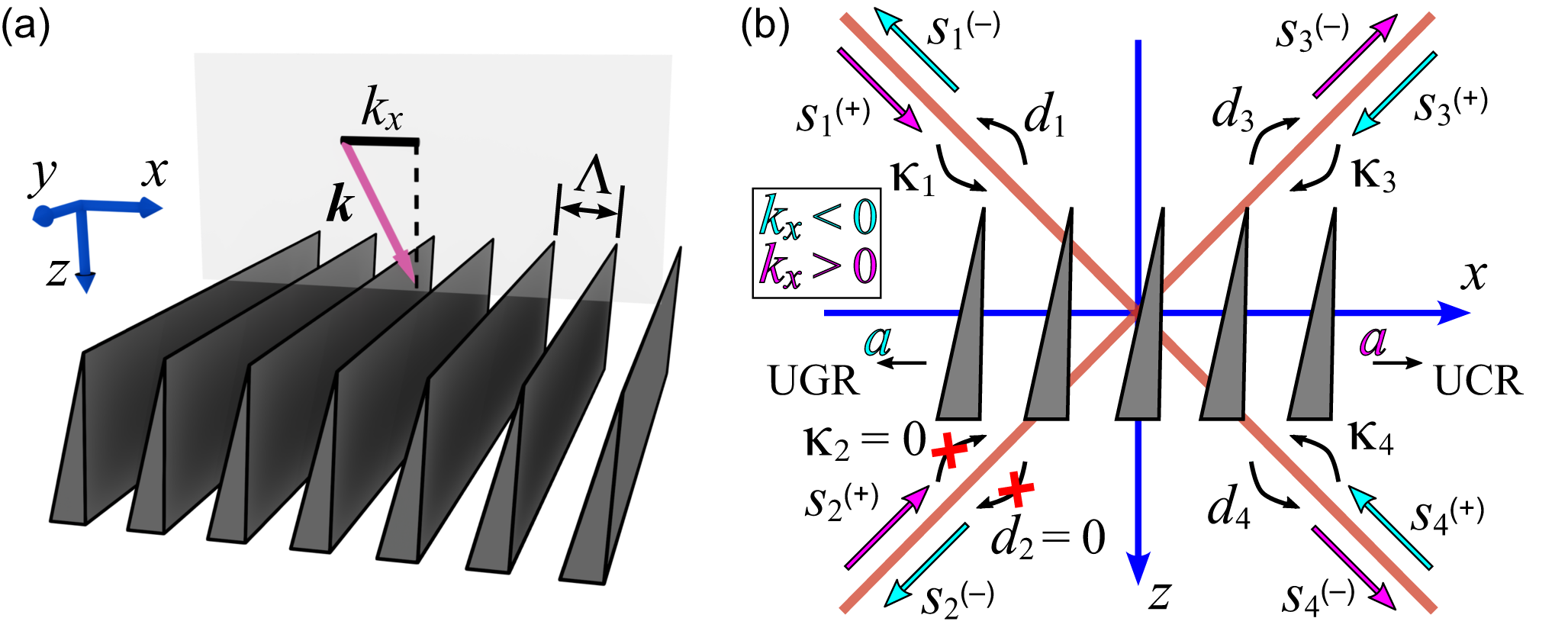}
    \caption{(a) Acoustic metasurface in the form of a ruled grating with the period $\Lambda$. The $xz$-plane of incidence is shaded in gray. The magenta arrow shows the wave vector $\bm k$ of the incident wave. The tangential component $k_x$ of the wave vector is indicated by the black segment. (b) Schematic of the TCMT model.
    The magenta and cyan arrows correspond to waves with positive ($k_x > 0$) and negative ($k_x < 0$) tangential components, respectively.
    The directions of the subscript arrows $a_{\smr, \sml}$ correspond to the mode propagation direction with
respect to the $x$-axis.}
    \label{Fig_1}
\end{figure}

In the case of an asymmetric metasurface, a UGR is
dual to a counter-propagating mode of a peculiar type --- a UCR, which radiates to both sides of the structure but, nonetheless, exhibits only single-sided coupling
with incident waves~\cite{maksimov2025temporal,pankin2025uniguided}.
When the in-plane mirror symmetry or inversion symmetry of the metasurface unit cell is restored, the counter-propagating modes have both properties, being simultaneously UGR and UCR~\cite{maksimov2025temporal}.

As a general wave phenomenon, the UGR concept was extended from photonics to acoustics in 2024 by Li et al.~\cite{li2024all}.
UGRs in acoustic metasurfaces with inversion symmetry of the unit cell have been proposed for intensity-flattened phase shifting~\cite{li2024all}, perfect sound absorption~\cite{li2025perfect}, and enhanced Goos-H\"anchen shifts~\cite{song2026large}.
We also note that UGRs can be implemented not only in metasurfaces but also in anisotropic 1D layered systems
in both photonics~\cite{mukherjee2021unidirectional, pankin2025uniguided} and acoustics~\cite{zhang2026anisotropy}.

In this Letter, we demonstrate UGR and UCR modes in an asymmetric acoustic metasurface.
The numerical calculations are confirmed by the temporal coupled-mode theory (TCMT).
By exploiting UGR/UCR modes, we demonstrate intensity-flattened phase shifting with an arbitrary transmittance in the lossless limit and perfect acoustic absorption only in the scattering channel coupled to the UCR.

\begin{figure}[tbp]
    \centering    \includegraphics[width=1\linewidth]{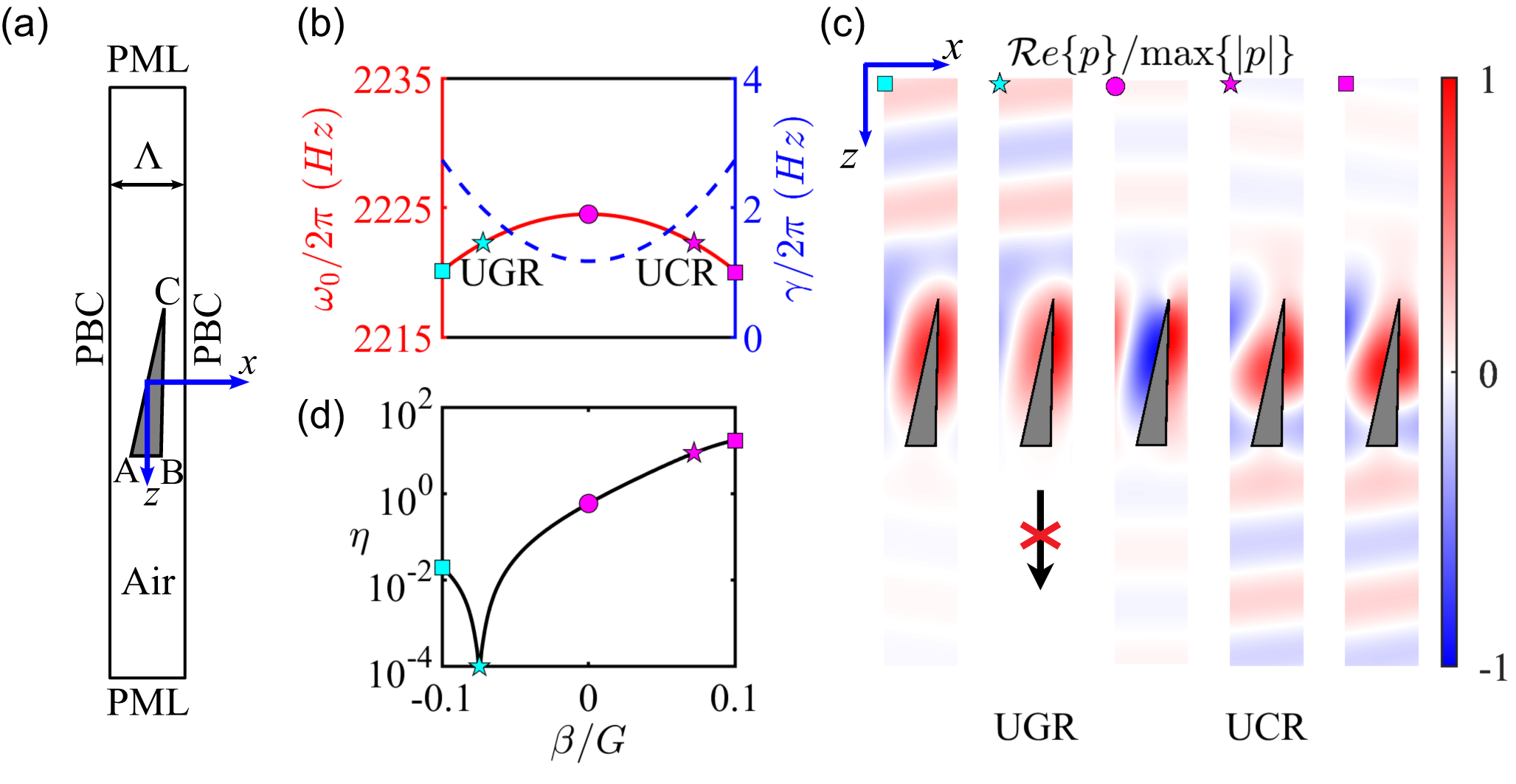}
    \caption{(a) The unit cell of the grating. The coordinates of the vertices in the $xz$-plane are A: (-2, 10) cm, B: (2, 10) cm, C: (2.4, -10) cm. Bloch-Floquet periodic boundary conditions (PBCs) are applied at the left and right boundaries of the unit cell. Perfectly matched layers (PMLs) are applied at the top and bottom boundaries. The period is $\Lambda = 10$ cm. The PMLs are placed at $z_t = -40$ cm and $z_b = 40$ cm. The panel (b) plots the natural frequency $\omega_0(\beta)$ and the radiative decay rate $\gamma_r(\beta)$. (c) Eigenmode pressure-field $p(x,z)$ profiles at the resonant frequencies $\omega_r$ marked in (b). (d) The radiation power ratio $\eta(\beta) = P_b/P_t$ is defined as $\eta(\beta) = 
\int |p(x,z_{b})|^2 \, dx/ \int |p(x,z_{t})|^2 \, dx$ with integration over the unit cell at the PML interfaces.
Since the substrate and superstrate are made of air with density $\rho_0$, the common integral prefactor $|k_z|/2\rho_0\omega$  for powers $P_b$ and $P_t$ are omitted from the fraction.}
\label{Fig_2}
\end{figure}

The system under scrutiny is a 2D acoustic metasurface made of rigid triangular ribs with hard boundary walls, invariant along the $y$-axis and periodic along the $x$-axis, as shown in Fig.~\ref{Fig_1}(a).
In the far field, the density is constant; thus, the monochromatic solutions are plane waves
\begin{equation}\label{11}
p(x,z)\propto e^{\pm ik_x x\pm ik_z z}, \ \ \frac{\omega^2}{c^2}=k_x^2+k_z^2,
\end{equation}
where $c = 343.2$ m$\cdot$s$^{-1}$ is the sound speed in air at a temperature of 293.15 K, and $k_{x,z}$ are the corresponding components of the wave vector. Since the metasurface is periodic, the solution has to comply with the Bloch theorem
\begin{align}\label{12}
 p(x,z)=p^{\scs(0)}(x,z)e^{i\beta x}, 
\ \ \beta\Lambda\in(-\pi, \pi],
\end{align}
where $p^{\scs(0)}(x,z)$ is a periodic function of $x$ with period $\Lambda$, and $\beta$ is the Bloch propagation constant. The conservation of quasi-momentum requires
$k_x=\beta+{2\pi m}/{\Lambda}$ where the index $m \in \mathbb{Z}$ is known as the diffraction order. 
The metasurface operates in the zeroth-order, or subwavelength regime, for which $m = 0$ and $\beta = k_x$, i.e. only the four zeroth-order channels are open, see Fig.~\ref{Fig_1}(b).
The eigenvalue and scattering problems were solved using the finite-element method (FEM) using the Pressure Acoustics, Frequency Domain interface in COMSOL Multiphysics.
The unit cell used in the simulations is shown in Fig.~\ref{Fig_2}(a).

The band structure $\omega_r(\beta)$ in the lossless case is shown in Fig.~\ref{Fig_2}(b), where  $G = 2\pi/\Lambda$ is the reciprocal lattice vector.
While the band structure is independent of the sign of the Bloch wavenumber $\beta$, the eigenmode profiles depend on its sign, see Fig.~\ref{Fig_2}(c). 
It can be seen that at $\beta \approx -0.072G$, a UGR occurs, since this resonance does not radiate in the bottom half-space.
The ratio between the power radiated by the resonance into the bottom ($P_b$) and top ($P_t$) half-spaces $\eta(\beta) = P_b/P_t$ also confirms the presence of the UGR, approaching zero at $\beta \approx -0.072G$, see Fig.~\ref{Fig_2}(d).
To realize a UGR, the width and height of the triangle were fixed, while the position of the top vertex was optimized to minimize $\eta$, which appears to be $\approx 10^{-4}$ in our simulation.
The convergence of $\eta$ with respect to mesh size and outgoing boundary positions is discussed in Supplemental Material ~\cite{suppl}.
As shown in Ref.~\cite{maksimov2025temporal}, in a metasurface with an asymmetric unit cell, the counter-propagating mode with the wavenumber $\beta \approx 0.072G$ is the UCR; see Fig.~\ref{Fig_2}(c).

\begin{figure}[tbp]
    \centering
    \includegraphics[width=\linewidth]{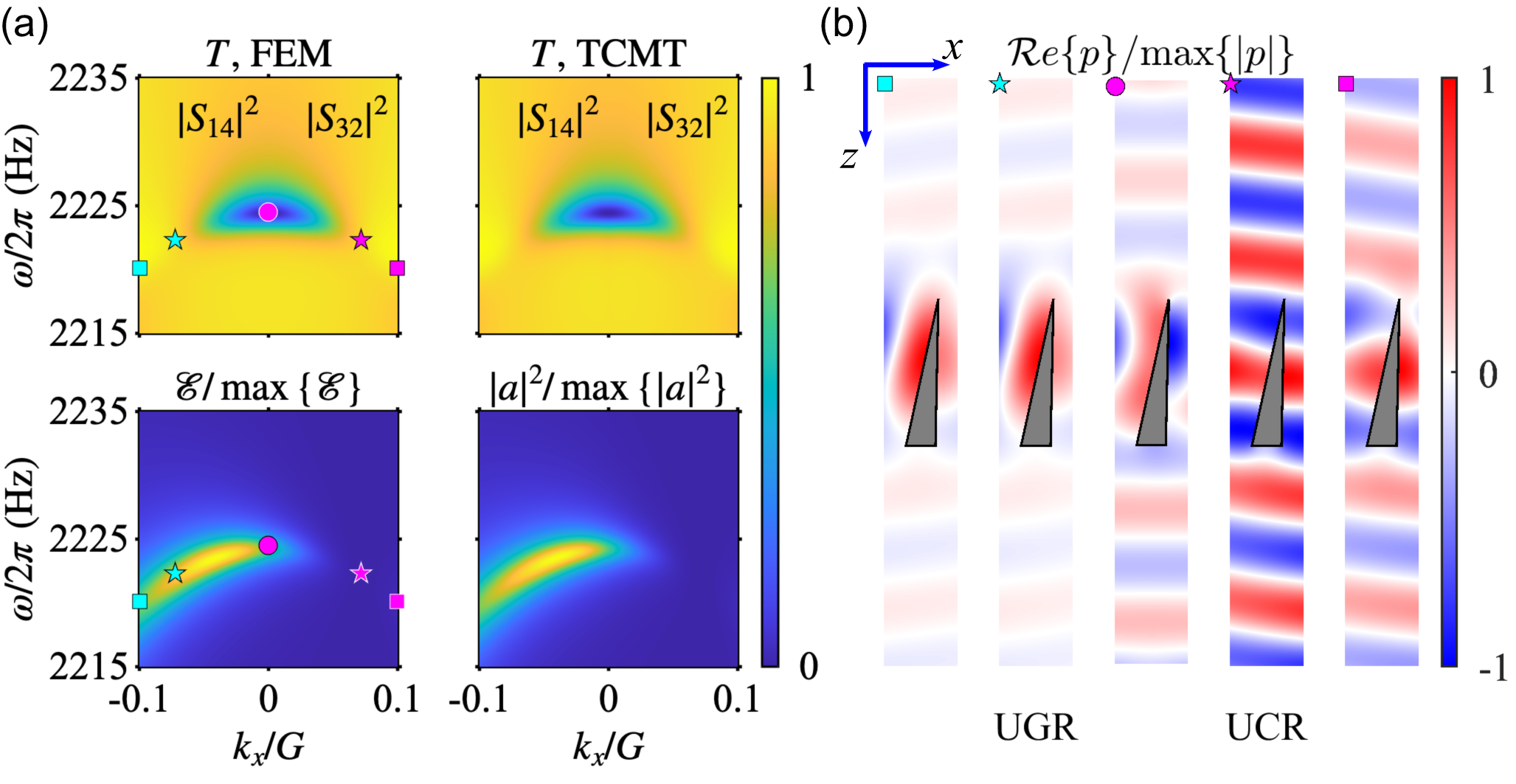}
    \caption{(a) Transmittance $T(k_x,\omega)$, calculated by the FEM and TCMT methods in the lossless case $\gamma_a = 0$. 
    The integrated pressure field intensity is calculated as $\cal E$$= \int_{-z_E}^{z_E} \int_{-\Lambda/2}^{\Lambda/2} |p(x,z)|^2 \, dx dz$, where $p(x,z)$ is the scattered field, $z_E = 20$ cm.
    (b) Total pressure-field $p(x,z)$ profiles at the resonant frequencies $\omega/2\pi = \omega_0/2\pi$ marked in (a).
    The panel corresponding to the UCR wavenumber $k_x = 0.072G$ (marked with the magenta pentagram) shows the incident field without excitation of the resonant mode.}
    \label{Fig_3}
\end{figure}

To confirm the unidirectional coupling of the UCR, we consider transmittance spectra and scattered fields in the lossless case for sound waves incident from the bottom side, i.e., the 2nd ($k_x>0$) and 4th ($k_x<0$) channels, see Fig.~\ref{Fig_1}(b).
The transmittance spectra exhibit a Fano resonance feature corresponding to resonant excitation, see Fig.~\ref{Fig_3}(a).
At $k_x \approx \mp 0.072G$ corresponding to the UGR/UCR, the resonant feature disappears, and  the transmittance is flat for both bottom-side inputs at these two points~\cite{li2024all}.
While the band structure is independent of the sign of the wavenumber $k_x$, the scattered-field profiles are dependent, see Fig.~\ref{Fig_3}(b).
It can be seen that in contrast to the UGR, the UCR is not excited by bottom-side incidence through the 2nd channel.
This fact is also confirmed by the map of the integrated scattered-pressure-field intensity $\cal E$$(k_x,\omega)$, see Fig.~\ref{Fig_3}(a).
Notably, the maximum of the resonant stored energy does not coincide with the UGR point, since its position is governed by the competition between the excitation efficiency and the radiative decay rate, rather than by the UGR condition alone, as discussed in Supplemental Material~\cite{suppl}. 

The spectral features of the UGR/UCR modes can be explained within the single-resonance TCMT framework derived in Ref.~\cite{maksimov2025temporal}, which we generalize here by introducing absorption losses
\begin{align}
\label{TCMT}
& \frac{d}{dt}a_{\smr, \sml} =
-\left(i\omega_0 + \gamma_r + \gamma_a \right)
a_{\smr, \sml} + {\kappa}_{\smr, \sml}^{\intercal}
{\bf s}_{\smr, \sml}^{\smallp}, \nonumber \\
&
{\bf s}_{\smr, \sml}^{\smallm} 
=\widehat{C}_{2\times2}
{\bf s}_{\smr, \sml}^{\smallp} + {\bf d}_{\smr, \sml} a_{\smr, \sml},
\end{align}
where the total decay rate $\gamma = \gamma_r + \gamma_a$ is the sum of the radiative $\gamma_r$ and absorption $\gamma_a$  decay rates.
By appropriately choosing the reference planes in each scattering channel~\cite{maksimov2025temporal}, the matrix describing lossless nonresonant scattering $\widehat{C}_{2\times2}$ can be written in the following form, 
\begin{equation}
\label{C}
   \widehat{C}_{2\times2}=
\left(
\begin{array}{cc}
\rho & i\tau \\
i\tau& \rho
\end{array}
\right).
\end{equation}
The vectors of coupling $\kappa$ and decoupling ${\bf d}$ coefficients are written as
\begin{equation}
{\kappa}_{\smr}=
\left(
\begin{array}{c}
\varkappa_{1} \\
\varkappa_{2}
\end{array}
\right),
\;
{\bf d}_{\smr}=
\left(
\begin{array}{c}
d_{3} \\
d_{4}
\end{array}
\right), \;{\kappa}_{\sml}=
\left(
\begin{array}{c}
\varkappa_{3} \\
\varkappa_{4}
\end{array}
\right), \;
{\bf d}_{\sml}=
\left(
\begin{array}{c}
d_{1} \\
d_{2}
\end{array}
\right),
\end{equation}
${\bf s}^{\smallp}$ and ${\bf s}^{\smallm}$ are vectors of incoming and outgoing wave amplitudes in the scattering channels, respectively, see Fig.~\ref{Fig_1}(b)
\begin{equation}
{\bf s}_{\smr}^{\smallp}=\!
\left(\!\begin{array}{c}
{ s}_{1}^{\smallp} \\
{ s}_{2}^{\smallp}
\end{array}
\!\right)\!, \
{\bf s}_{\smr}^{\smallm}=
\left(\!\begin{array}{c}
{ s}_{3}^{\smallm} \\
{ s}_{4}^{\smallm}
\end{array}
\!\right), \
{\bf s}_{\sml}^{\smallm}=
\left(\!\begin{array}{c}
{ s}_{1}^{\smallm} \\
{ s}_{2}^{\smallm}
\end{array}\!
\right), \
{\bf s}_{\sml}^{\smallp}=
\left(\!\begin{array}{c}
{ s}_{3}^{\smallp} \\
{ s}_{4}^{\smallp}
\end{array}\!
\right).
\end{equation}
The amplitudes are normalized such that $|a|^2$ represents the energy stored in the resonant mode, while $|s^{(\pm)}|^2$ represents the incoming or outgoing acoustic power. The total decay rate $\gamma$ describes the decay of the modal amplitude, $a(t)\propto e^{-\gamma t}$, such that the stored energy decays as $|a(t)|^2\propto e^{-2\gamma t}$.
 
The coupling vectors are related to the amplitude coefficients of non-resonant reflection $\rho = \sqrt{1 - \tau^2}$ and transmission $\tau\in[-1, 1]$ through
\begin{align}
\label{kappa_final}
{\kappa}_{\smr,\sml}=\!
\!\sqrt{\frac{\gamma_r}{1+\rho}}
\!\left(\!
\begin{array}{c}
\tau\cos\alpha~e^{\mp i\delta/2} -i(1+\rho)\sin\alpha~e^{\pm i\delta/2} \\
-i(1+\rho)\cos\alpha~e^{\mp i\delta/2}+\tau\sin\alpha~e^{\pm i\delta/2}
\end{array}
\!\right)\!,
\end{align}
where $\alpha\in[-\pi/2, \pi/2]$ and $\delta\in[0, \pi)$ are TCMT fitting parameters. The upper signs correspond to $\kappa_{\smr}$ and the lower signs correspond to $\kappa_{\sml}$.
Lorentz reciprocity imposes the following relations between the vectors of coupling and decoupling coefficients
\begin{equation}\label{coup_decoup}
{\bf d}_{\sml}={ \kappa}_{\smr}, \ {\bf d}_{\smr}={\kappa}_{\sml}.
\end{equation}
Direct substitution of Eq.~\eqref{kappa_final} shows that decoupling coefficients satisfy the energy conservation law ${\bf d}_{\smr,\sml}^{\dagger} {\bf d}_{\smr,\sml} = 2\gamma_r$ in the absence of material losses.
Due to momentum conservation, the $4\!\times\!4$ scattering matrix $\widehat{S}$ should have a block form \begin{equation}
\left(\begin{array}{c}
{\bf s}_{\sml}^{\smallm} \\
{\bf s}_{\smr}^{\smallm}
\end{array}\right)=
\widehat{S}
\left(
\begin{array}{c}
{\bf s}_{\smr}^{\smallp} \\
{\bf s}_{\sml}^{\smallp}
\end{array}
\right), \; \widehat{S} = \left(
\begin{array}{cc}
0 & \widehat{S}_{2\times2} \\
\widehat{S}_{2\times2}^{\intercal} & 0
\end{array}
\right).
\end{equation}
\begin{figure}[tbp]
    \centering
    \includegraphics[width=\linewidth]{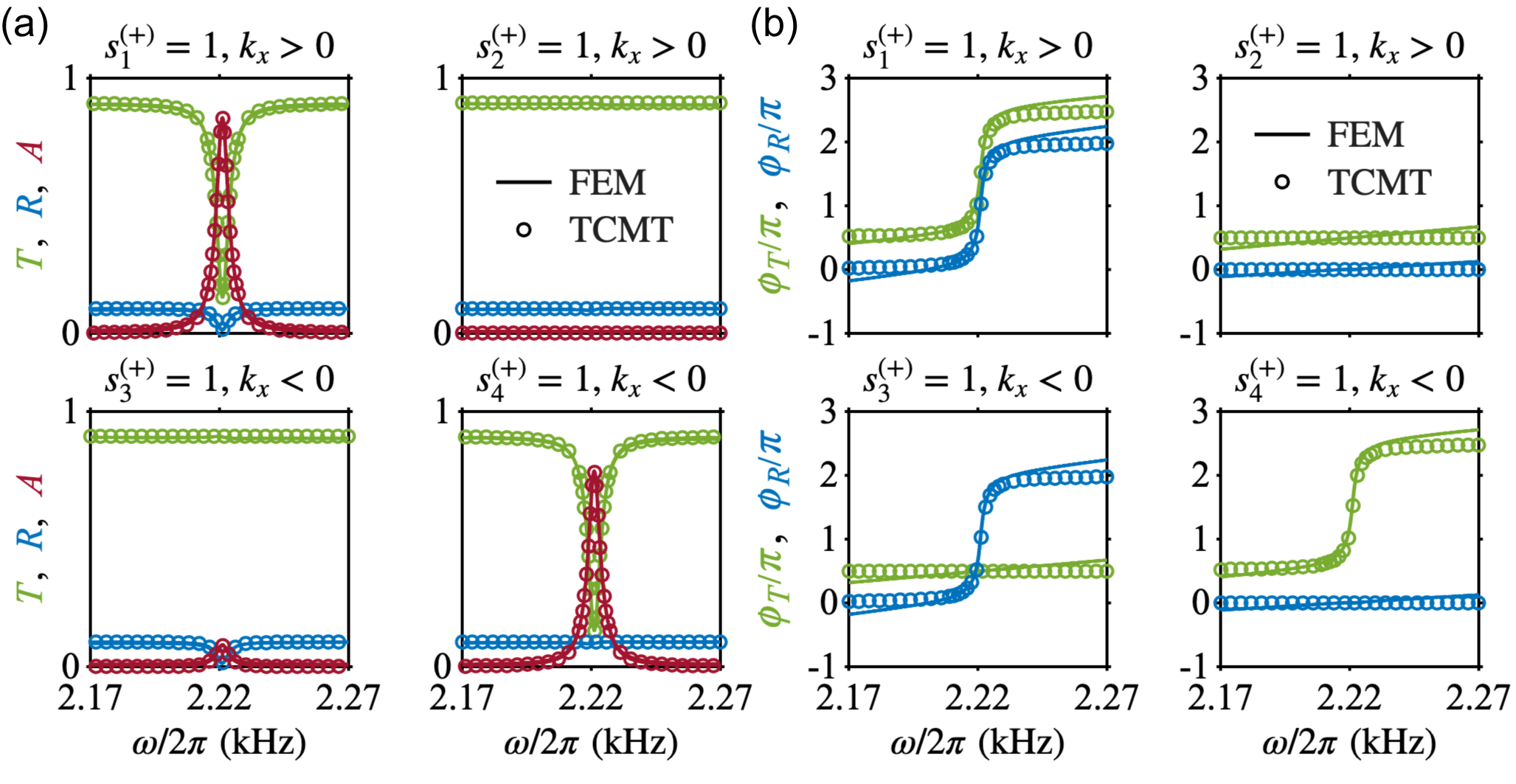}
    \caption{(a) Scattering spectra $T$ (green), $R$ (blue), $A$ (red) for UGR/UCR at $k_x = \mp 0.072G$.
    (b) Unwrapped transmission (green) and reflection (blue) phase response $\phi = \operatorname{unwrap}[\arg\{S_{jj^{\prime}}\}]$ of the asymmetric metasurface for UGR/UCR at $k_x = \mp 0.072G$. 
The principal argument is taken in the interval
$(-\pi,\pi]$, and phase unwrapping is performed along
increasing frequency by adding integer multiples of
$2\pi$ to remove branch discontinuities.
The initial phase at the lowest plotted frequency
is chosen on the principal branch.
 }
    \label{Fig_4}
\end{figure}
By applying a time-harmonic substitution $e^{-i\omega t}$ for all
time-dependent amplitudes in Eq.~\eqref{TCMT}, one obtains an expression for the scattering coefficients
\begin{equation}
\label{Smat}
S_{jj^{\prime}} = C_{jj^{\prime}} + \frac{d_j \kappa_{j^{\prime}}}{i(\omega_0 - \omega) + \gamma_r + \gamma_a},
\end{equation}
where $j = 1,2 (3,4)$, $j^{\prime} = 3,4 (1,2)$, and
\begin{equation}
 \widehat{C}  =
\left(
\begin{array}{cc}
0 & \widehat{C}_{2\times2} \\
\widehat{C}_{2\times2}^{\intercal} & 0
\end{array}
\right).
\end{equation}

The absence of UGR radiation into the second channel imposes the condition $d_2=0$, while the reciprocity relation in Eq.~\eqref{coup_decoup} leads to the UCR condition $\kappa_2 = 0$. Substitution of this condition into Eq.~\eqref{kappa_final} yields the following constraints on the TCMT parameters
\begin{equation}
\label{UGR_cond1}
\tau = 0, \quad\; \alpha = \pm \frac{\pi}{2}, \quad \text{and} \quad
\delta=\frac{\pi}{2}, \quad
\tan\alpha=\frac{1+\rho}{\tau}.
\end{equation}
These relations therefore define the conditions under which radiation into the second channel is completely suppressed.
The parameters \(\tau (|k_x|)\), \(\alpha (|k_x|)\) and \(\delta(|k_x|)\) can be approximated by polynomial interpolation.
In the case discussed in this Letter, $\tau \neq 0$, thus, Eq.~\eqref{UGR_cond1} was used at the UGR/UCR point $|k_x| = 0.072G$ for interpolation, along with the condition at normal incidence $\delta(0) = 0$, see Supplemental Material~\cite{suppl}.
This allows the transmittance and stored energy $|a|^2$ to be evaluated as a function of the tangential wavenumber \(k_x\) and the frequency \(\omega\), see Fig.~\ref{Fig_3}(a).
The transmittance spectra calculated within the TCMT framework in the lossless case $\gamma_a = 0$ are in agreement with those obtained by FEM, demonstrating the consistency of the analytical model with the numerical results.

The scattering coefficients in Eq.~\eqref{Smat} 
under the UGR/UCR conditions in Eq.~\eqref{UGR_cond1} can be written as

\begin{align}
\label{SUGR}
& S_{41}=S_{14}=i\tau \left[\frac{i(\omega_0-\omega)-\gamma_r+\gamma_a}{i(\omega_0-\omega)+\gamma_r+\gamma_a} \right], \nonumber \\
& S_{31}=S_{13}=\rho \left[\frac{i(\omega_0-\omega)-\gamma_r+\gamma_a}{i(\omega_0-\omega)+\gamma_r+\gamma_a} \right], \nonumber \\
& S_{23}=S_{32}=i\tau ,
\quad S_{24}=S_{42}=\rho,
\end{align}
which yields the corresponding spectral coefficients, namely transmittance $T$ and reflectance $R$ 
\begin{align}
\label{S_coef}
& T: \;  |S_{41}|^2=|S_{14}|^2
=\tau^2\left[1-\frac{4\gamma_r\gamma_a}
{(\omega-\omega_0)^2+(\gamma_r+\gamma_a)^2}\right], \nonumber\\
& R: \; |S_{31}|^2=|S_{13}|^2
=\rho^2\left[1-\frac{4\gamma_r\gamma_a}
{(\omega-\omega_0)^2+(\gamma_r+\gamma_a)^2}\right], \nonumber\\
& T: \; |S_{23}|^2=|S_{32}|^2=\tau^2, \quad
R: \; |S_{24}|^2=|S_{42}|^2=\rho^2.
\end{align}
Equations~\eqref{SUGR} are applicable to the case of UGR in both photonics and acoustics, with in-plane mirror symmetry of the unit cell, for which $\tau = 0, \; \rho = 1$~\cite{Zhang2022} and inversion symmetry, for which $\tau = 1, \; \rho = 0$ at UGR~\cite{Zhang2022, li2024all}.
It can be seen that the intensity-flattened phase shifting, i.e. frequency-independent resonant spectral coefficients, reported in~\cite{Zhang2022, li2024all} can be achieved only in the absence of absorption decay rate $\gamma_a = 0$, see Supplemental Material~\cite{suppl}.

Fig.~\ref{Fig_4} presents the spectral coefficients and their phase responses for the lossy case.
In COMSOL Multiphysics the losses were taken into account via the Thermoviscous Boundary Layer Impedance (TBLI), see physical parameters in Supplemental Material~\cite{suppl}.
The resulting spectral response clearly exhibits the resonant features in the spectral coefficients shown in Fig.~\ref{Fig_4}(a), accompanied by the resonant
phase shifts,   see Fig.~\ref{Fig_4}(b).
The phase response and spectra predicted by TCMT closely matches those obtained by FEM.
The difference between two methods is caused by TCMT assumptions of frequency independent and lossless direct process, i.e. $\widehat{C} \neq \widehat{C}(\omega)$ and $\widehat{C}^{\dagger}\widehat{C} = \widehat{\mathbb{I}}$.
It can be seen that FEM phase response has a slope, and absorption in direct process, i.e. $\tau^2 + \rho^2 \neq 1$  
.
Nevertheless, the depth of the resonance dip can be reduced by reducing the value of $\gamma_a$.
In the case of acoustics, thermoviscous losses are frequency-dependent $\gamma_a \propto \sqrt{\omega}$, and can be reduced by scaling the structure size, see corresponding spectra in Supplemental Material~\cite{suppl}.

\begin{figure}[tbp]
    \centering
    \includegraphics[width=\linewidth]{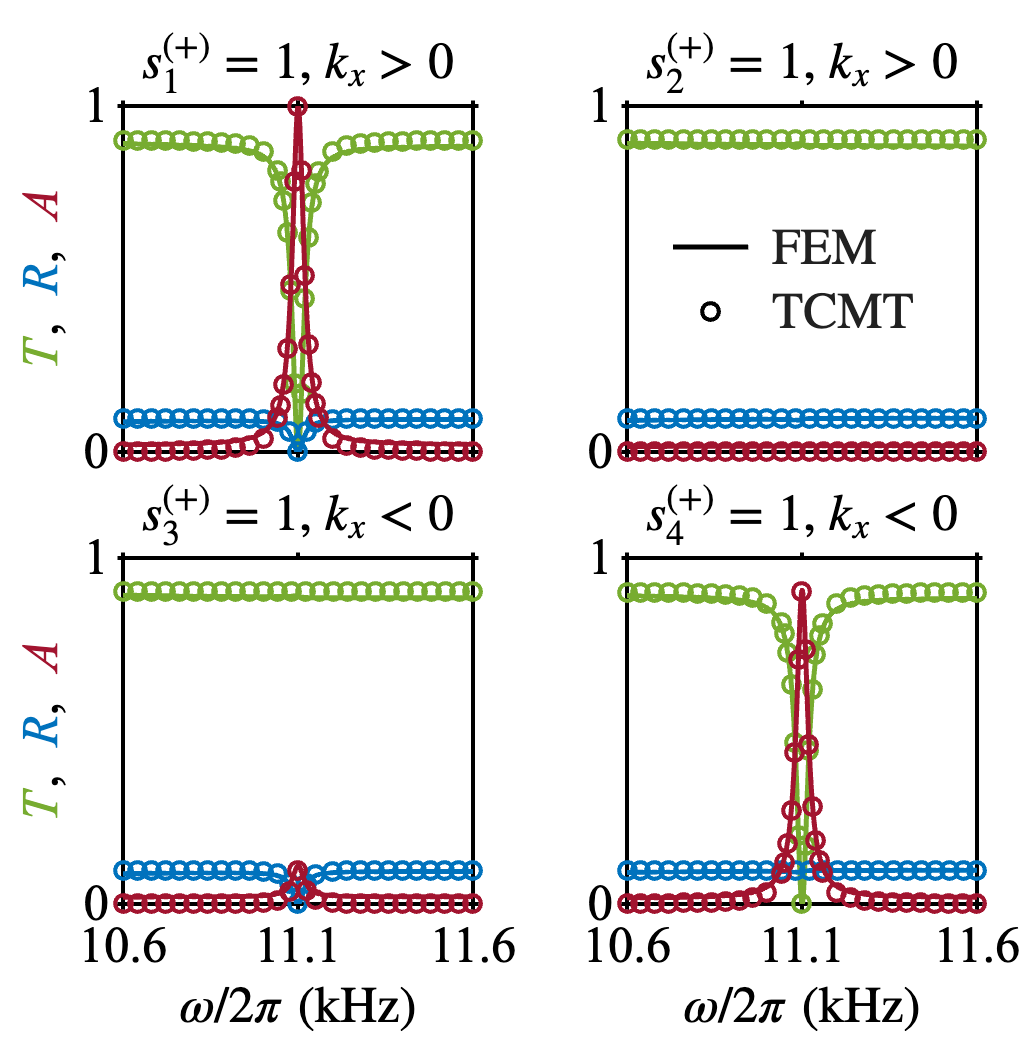}
    \caption{(a) Scattering spectra $T$ (green), $R$ (blue), $A$ (red) for UGR/UCR at $k_x = \mp 0.072G$.
    Critical coupling is implemented at $\omega/2\pi = \omega_0/2\pi = 11.1$ kHz for the scaled structure with decay rates $\gamma_r/2\pi = \gamma_a/2\pi = 9.93$ Hz. PSA is achieved for sound incident from the 1st scattering channel. 
    The period of the structure is $\Lambda = $ 2 cm. The coordinates of the vertices in the $xz$-plane are A: (-0.4, 2) cm, B: (0.4, 2) cm, C: (0.48, -2) cm.}
    \label{Fig_5}
\end{figure}

The extended TCMT theory enables the investigation of perfect sound absorption (PSA) at UGR/UCR within the TCMT formalism.
The absorption coefficient \(A_{j^{\prime}}\) for a sound wave incident through the \(j^{\prime}\)-th scattering channel can be defined using Eq.~\eqref{S_coef} as
\begin{align}
& A_{1} \equiv 1 -|S_{41}|^2-|S_{31}|^2= \frac{4\gamma_r\gamma_a}{(\omega - \omega_0)^2 + (\gamma_r + \gamma_a)^2},
\nonumber\\
& A_{2} \equiv 1 -|S_{32}|^2-|S_{42}|^2 = 0,
\nonumber\\
&A_{3} \equiv 1 -|S_{23}|^2-|S_{13}|^2= \frac{4\rho^2\gamma_r\gamma_a}{(\omega - \omega_0)^2 + (\gamma_r + \gamma_a)^2},
\nonumber\\
&A_{4} \equiv 1 -|S_{14}|^2-|S_{24}|^2 = \frac{4\tau^2\gamma_r\gamma_a}{(\omega - \omega_0)^2 + (\gamma_r + \gamma_a)^2}.
\end{align}

These equations demonstrate that the resonant absorptances $A_1$, $A_3$ and $A_4$ reach their maxima at the critical-coupling conditions \(\omega=\omega_0\) and \(\gamma_r=\gamma_a\)~\cite{wu2021quasi}.
Under these conditions, the TCMT model predicts PSA for excitation through the 1st channel
\begin{equation}
    \label{PSA}
    A_1 = 1, \; A_2 = 0, \; A_3 = \rho^2, \; A_4 = \tau^2.
\end{equation}
The 1st and 2nd channels belong to the UCR momentum sector, with excitation possible only through the 1st channel, while UGR can be excited through the 3rd and 4th channels, see Fig.~\ref{Fig_1}(b).
Thus, in the case of single-channel incidence, the critical-coupling conditions result in a highly asymmetric absorption, with perfect absorption occurring for excitation of the UCR from the only side from which it can be excited.
This effect is in contrast to the previously reported perfect absorption enabled by a UGR in photonics~\cite{Xu2023a} and acoustics~\cite{li2025perfect}, for metasurfaces with unit-cell inversion symmetry.
Since the two UGR modes are also UCRs, in the above-mentioned papers, perfect absorption is possible for single-channel incidence from two sides of the metasurface, namely, for channels with different signs of tangential momentum $k_x$, i.e. 1st and 4th or 2nd and 3rd scattering channel pairs.
It should be mentioned that the asymmetric absorption response in our case is compatible with reciprocity, i.e. $S_{jj^{\prime}} = S_{j^{\prime}j}$.

To verify PSA numerically, we scaled the system to a higher frequency range, to enlarge the absorption decay rate and achieve the critical coupling condition $\gamma_r = \gamma_a$.
It can be shown that under the coordinate transformation
$\mathbf{r}(N)=N\mathbf{r}$ for a given metasurface design, the complex eigenfrequency $\omega_r = \omega_0 - i(\gamma_r + \gamma_a)$  is scaled with the geometrical scaling factor $N$ with the following phenomenological rule, see Supplemental Material~\cite{suppl}
\begin{equation}
\omega_r(N)
=
\frac{\omega_0}{N}
-i\left(
\frac{\gamma_r}{N}
+
\frac{\gamma_a}{N^{3/2}}
\right),
\end{equation}
so the scaling factor for PSA can be calculated via
\begin{equation}
\label{Scaling}
\gamma_r(N)=\gamma_a(N)
\quad\Longrightarrow\quad
\frac{\gamma_r}{N}=\frac{\gamma_a}{N^{3/2}}
\quad\Longrightarrow\quad
N=\left(\frac{\gamma_a}{\gamma_r}\right)^2.
\end{equation}
In our case, at $\omega_0/2\pi = 2222$ Hz, the corresponding values are $\gamma_r/2\pi = 1.98$ Hz and $\gamma_a/2\pi = 0.87$ Hz.
Using Eq.~\eqref{Scaling}, we can obtain $N \approx 0.2$.

The spectra for the correspondingly scaled structure are shown in Fig.~\ref{Fig_5}.
Agreement with the TCMT prediction provides a clear physical interpretation of this unidirectional perfect absorption mechanism.
The residual discrepancy between FEM and TCMT can be seen in the background absorption, which yields $A_2 = 0.016$ in FEM, while $A_2 = 0$ in TCMT.
As was mentioned above, this is caused due to the assumption of lossless background scattering in TCMT.
It should be mentioned that thermoviscous losses in a real experiment can be enhanced not only by scaling, but also by introducing a thin layer of absorbing foam at the surface of the ribs \cite{wu2026experimental}.
It should be noted that in our asymmetric case, $\tau$ can be arbitrary.
This is distinct from the previously reported perfect absorption in photonics and acoustics in the metasurfaces with the inversion symmetry of the unit cell, for which $\tau = 1$~\cite{Xu2023a,li2025perfect}.

In conclusion, this work demonstrates a distinct pair of counter-propagating uniguided and unicoupled resonances in an asymmetric acoustic grating.
Unlike the uniguided resonance, the unicoupled resonance couples to outgoing waves on both sides of the metasurface but, nonetheless, exhibits only single-sided coupling
with incident waves.
Temporal coupled-mode theory, which accounts for asymmetric coupling, has been generalized to account for absorption losses.
Within the developed theoretical model, we demonstrate intensity-flattened phase shifting with arbitrary transmittance in the lossless limit.
The model predicts perfect sound absorption at critical coupling, reproduced numerically by a rigorous finite-element method.
In contrast to metasurfaces with inversion symmetry of the unit cell, in the asymmetric case, perfect absorption is allowed only for the scattering channel corresponding to excitation of the unicoupled resonance.
We speculate that the proposed uniguided and unicoupled resonances in asymmetric metasurfaces pave the way to engineering acoustic devices with asymmetric coupling, including couplers, sound emitters, and ventilated absorbers.

\begin{acknowledgments}
This work was funded by the Russian Science Foundation grant № 26-72-10076. The work of P.S.P. on developing the TCMT theory was financially supported by the ITMO Fellowship Program.
\end{acknowledgments}

\bibliography{References}

% \appendix

\end{document}